\documentclass[aps,prd,showpacs,
amsmath,amssymb,floatfix,nofootinbib,english]
{revtex4-1} 
\usepackage[T1]{fontenc}
\usepackage[latin9]{inputenc}
\usepackage{textcomp}
\usepackage{amssymb}
\usepackage{epsfig}
\usepackage{graphicx}
\usepackage{esint}
\usepackage{comment}
\usepackage{color}
\PassOptionsToPackage{normalem}{ulem}
\usepackage{ulem}
\usepackage{babel}
\usepackage{slashed}
\newcommand{\be}{\begin{equation}}
\newcommand{\ee}{\end{equation}}
\newcommand{\bea}{\begin{eqnarray}}
\newcommand{\eea}{\end{eqnarray}}

\newcommand{\ba}{\begin{array}}
\newcommand{\ea}{\end{array}}
\newcommand{\bd}{\begin{displaymath}}
\newcommand{\ed}{\end{displaymath}}

\def\gev{{\rm \,Ge\kern-0.125em V}}
\def\tev{{\rm \,Te\kern-0.125em V}}

{\newcommand{\lsim}{\mbox{\raisebox{-.6ex}{~$\stackrel{<}{\sim}$~}}}
{\newcommand{\gsim}{\mbox{\raisebox{-.6ex}{~$\stackrel{>}{\sim}$~}}}

\def\a{\alpha}

\def\th13 {\theta_{13}}

\begin{document}

\title{Nonthermal $CP$ violation in soft leptogenesis}

\author{Rathin Adhikari}
\email{rathin@ctp-jamia.res.in}
\author{Arnab Dasgupta}
\email{arnab@ctp-jamia.res.in}
\affiliation{Centre for Theoretical Physics, Jamia Millia Islamia (Central University), Jamia Nagar, New Delhi 110025, India}
\author{Chee Sheng Fong}
\email{fong@if.usp.br}
\affiliation{Instituto de F\'{\i}sica, 
Universidade de S\~ao Paulo, C.\ P.\ 66.318, 05315-970 S\~ao Paulo, Brazil}
\author{Raghavan Rangarajan}
\email{raghavan@prl.res.in}
\affiliation{Theoretical Physics Division, Physical Research Laboratory,
Navrangpura, Ahmedabad 380009, India}

\date{\today}

\begin{abstract}
Soft leptogenesis is a mechanism which generates 
the matter-antimatter asymmetry of the Universe via the 
out-of-equilibrium decays of heavy sneutrinos 
in which soft supersymmetry breaking terms
play two important roles: they provide the required $CP$ violation 
and give rise to the mass splitting between otherwise 
degenerate sneutrino mass eigenstates within a single generation. 
This mechanism is interesting because it can be successful at the
lower temperature regime $T \lesssim 10^9$ GeV in which 
the conflict with the overproduction of gravitinos can possibly be avoided.
In earlier works the leading CP violation is found to be
nonzero only if finite temperature effects are included.
By considering generic soft trilinear couplings, 
we find two interesting consequences: 
(1) the leading $CP$ violation can be nonzero even at zero temperature
realizing nonthermal CP violation, and 
(2) the $CP$ violation is sufficient even far away from the resonant regime 
allowing soft supersymmetry breaking parameters to assume natural 
values at around the TeV scale. We discuss phenomenological 
constraints on such scenarios and conclude that the contributions
to charged lepton flavor violating 
processes are close to the sensitivities of present and future experiments.
\end{abstract}


\maketitle

\section{Introduction}

Leptogenesis \cite{Fukugita:1986hr} is an attractive mechanism for generating the observed 
matter-antimatter asymmetry of the Universe wherein one first creates an asymmetry 
in the lepton sector which, in turn, induces an asymmetry in the baryon sector 
via anomalous $B+L$ violating interactions. 
In standard type-I seesaw supersymmetric leptogenesis \cite{Campbell:1992hd,
Covi:1996wh,Plumacher:1997ru,Fong:2010qh}
involving the out-of-equilibrium decays of heavy neutrinos and sneutrinos, 
the $CP$ violation required to generate the lepton number asymmetry
comes from the neutrino Yukawa couplings.  This scenario, with
hierarchical right-handed neutrinos (RHNs), faces a conflict as successful
leptogenesis requires the mass of the lightest RHN to be at 
least $10^9\gev$ \cite{Davidson:2002qv} while the 
simplest resolution of the gravitino problem 
\cite{Pagels:1981ke,Weinberg:1982zq}
requires
the reheating temperature after inflation to be less than $10^{6\text{--}9}\gev$
depending on the gravitino mass \cite{gravitino}.\footnote{See 
Refs. \cite{Allahverdi:2005fq,Allahverdi:2005mz,Rangarajan:2012wy} 
for another resolution of the gravitino problem due
to delayed thermalization of the Universe after inflation.}

One may avoid this conflict by incorporating new elements in leptogenesis.
In models of soft leptogenesis \cite{D'Ambrosio:2003wy,Grossman:2003jv} 
(for a recent review, see Ref.~\cite{Fong:2011yx})
$CP$ violation comes from soft supersymmetry (SUSY) breaking terms 
(here onwards we will simply refer to them as soft terms) 
with soft parameters assumed to be at the $m_{\rm SUSY} \sim$ TeV scale; i.e., 
we still hope SUSY is responsible for stabilizing the 
hierarchy between the weak and grand unification scales.
One interesting feature is that soft leptogenesis can 
proceed even with one generation of the RHN chiral superfield.\footnote{In a 
realistic model, we need at least two RHNs to accommodate neutrino oscillations.
Assuming RHNs to be hierarchical, soft leptogenesis only depends 
on the parameters related to the lightest RHN and decouples from
the parameters related to heavier RHNs.}
Essentially, the heavy sneutrino $\widetilde{N}$ and antisneutrino 
$\widetilde{N}^*$ from the same chiral supermultiplet 
will mix due to the presence of the soft terms. 
The decays of the mixed mass eigenstates
violate both $CP$ and lepton number and generate a matter-antimatter asymmetry. 
Although the $CP$ violation is suppressed by powers of $m_{\rm SUSY}/M \ll 1$ 
with $M$ the scale of the lightest RHN, the mass splitting 
between these otherwise degenerate sneutrino mass eigenstates
is also proportional to $m_{\rm SUSY}/M$. Crucially, this small splitting also 
results in enhancement of the $CP$ violation from mixing. 
Because of the suppression factor $m_{\rm SUSY}/M$ in the $CP$ violation, 
one cannot have very large $M$. Estimating the leading $CP$ parameter as 
$\epsilon \sim m_{\rm SUSY}/M$ and that successful leptogenesis generically 
requires $\epsilon \gtrsim 10^{-6}$, we obtain $M \lesssim 10^{9}$ GeV 
assuming $m_{\rm SUSY}$ at the TeV scale.
Hence soft leptogenesis occurs in the regime where the conflict with the 
bound on the reheating temperature from gravitino overproduction can be 
mitigated or even avoided.

In the original proposals of Refs. \cite{D'Ambrosio:2003wy,Grossman:2003jv},
the authors showed that in the scenario of $\widetilde{N}-\widetilde{N}^*$
mixing, the leading $CP$ violation in decays to fermions and scalars 
have opposite signs and cancel each other at the order 
${\cal O}\left(m_{{\rm SUSY}}/M\right)$ at zero temperature $T=0$.
They further showed that once finite temperature effects are taken into account, 
this cancellation is partially lifted,
i.e. one obtains an asymmetry proportional to a factor [$c_F(T)-c_B(T)$], 
where $c_{F,B}(T)$ are phase space and statistical 
factors associated with fermion and boson final states, and
where the contributions do not completely cancel each other at finite temperature.
Working under the assumption of 
proportionality of soft trilinear couplings $A_\alpha = A Y_\alpha$ where the $Y_\alpha$'s 
are the neutrino Yukawa couplings and $\alpha$ the lepton flavor index,
they showed that the resulting $CP$ violation is of the order of 
${\cal O}\left(m_{{\rm SUSY}}/M\right)$ at the \emph{resonance} 
which, however, requires an unconventionally small soft bilinear coupling
$B \ll m_{\rm SUSY}$. Away from the resonance, the $CP$ violation is of
${\cal O}\left(Y_\alpha^2\right)$ and, hence, 
too suppressed for successful leptogenesis.
On the other hand, assuming generic $A$ couplings, 
Ref.~\cite{Fong:2010zu} showed that successful leptogenesis 
can be obtained with $B \sim m_{\rm SUSY}$ away from the resonant regime.

Later in Ref.~\cite{Grossman:2004dz} it was argued that direct
$CP$ violation, i.e., from vertex corrections, due to gaugino exchange in the loop, 
survives at the order ${\cal O}\left(m_{{\rm SUSY}}^{2}/M^{2}\right)$ at $T=0$. 
Since the neutrino Yukawa coupling is replaced by the 
gauge coupling in the $CP$ violation parameter,
a large $CP$ violation can be obtained for $M$ at the TeV scale. 
Further study in Ref.~\cite{Fong:2009iu}, however, showed that in fact 
in this scenario, the cancellation still holds up 
to ${\cal O}\left(m_{{\rm SUSY}}^{2}/M^{2}\right)$ 
at $T=0$, and it was concluded that finite temperature effects 
are necessary to prevent the cancellation.
The cancellation is consistent with the result obtained 
in Ref.~\cite{Adhikari:2001yr} which states that 
to have a nonvanishing total $CP$ violation there should be lepton number
violation to the right of the ``cut'' in the loop diagram, and 
this requirement is not fulfilled in these cases. 
More recently, in Ref.~\cite{Garbrecht:2013iga} it was shown
that if finite temperature effects are taken into account \emph{consistently},
the cancellation of direct $CP$ violation from 
the gaugino contribution still holds even at $T\neq 0$.

In fact, in soft leptogenesis at finite temperature, the partial cancellation in the resulting 
lepton and slepton number density asymmetries sourced by $CP$ violation from mixing 
and the complete cancellation in the case of the gaugino vertex correction \cite{Garbrecht:2013iga} 
only hold under the assumption of equilibration between the chemical potentials 
of leptons and sleptons (superequilibration) which is valid below $T\lsim 10^8$ GeV for
$m_{\rm SUSY} \sim $ TeV  \cite{Fong:2010qh}.  As shown in Ref.~\cite{Fong:2010bv}, 
in the nonsuperequilibration regime, the partial cancellation between lepton and slepton 
number density asymmetries in the mixing scenario is avoided, resulting in an enhanced efficiency 
for soft leptogenesis. However, for reasons given later, 
we shall below consider mixing and vertex scenarios in the superequilibration regime 
(and also find a case where the lepton and slepton number density asymmetries
 do not partially cancel each other).
On the other hand, considering  $M \gtrsim 10^8$ GeV and $m_{\rm SUSY} \sim 1$ TeV, the 
$CP$ violating parameter from the gaugino contribution in the nonsuperequilibration regime is
$\epsilon \sim 10^{-1} m_{\rm SUSY}^2/M^2 \lesssim 10^{-11}$ and, hence, is too small for
successful leptogenesis.  Therefore processes involving gauginos
will not be considered further in this work.

In this article, we revisit soft leptogenesis by relaxing 
the assumption of the proportionality of the $A$ couplings.
In Sec. \ref{sec:lag}, we review the Lagrangian for soft leptogenesis 
with generic $A_\alpha$ terms and spell out the constraints 
from out-of-equilibrium decays of heavy sneutrinos 
and the cosmological bound on the sum of neutrino masses. 
In Sec. \ref{sec:CP_violation}, we obtain the $CP$ violating
parameter for both the self-energy corrections (mixing) and vertex corrections.
We show that generic $A_\alpha$ couplings give rise to two interesting consequences: 
(1) the leading $CP$ violation can be nonzero even when thermal corrections are neglected 
implying a \emph{nonthermal} $CP$ violation, and 
(2) the mixing $CP$ violation away from the resonance is of the order of 
${\cal O}(Y_\alpha)$ and, hence, can be large enough for leptogenesis. 
Because of the small mass splitting, the mixing $CP$ violation always dominates 
over the vertex $CP$ violation even far away from the resonant regime.
In Ref.~\cite{Kashti:2004vj}, it was shown that with $A_\alpha = A Y_\alpha$,
soft leptogenesis gives negligible contributions to the electric dipole moment
of charged leptons and charged lepton flavor violating processes.   
In Sec. \ref{sec:pheno}, we repeat the exercise and show that 
with generic $A_\alpha$ couplings, the contributions to charged lepton flavor 
violating processes are close to the sensitivities of present and future experiments. 
Finally, in Sec. \ref{sec:conclusions}, we conclude. 
This article is completed with two appendixes. 
In Appendix \ref{app:therm}, we discuss the inclusion of thermal effects 
under the assumption of decaying heavy sneutrinos at rest. 
In Appendix \ref{app:CP_asym}, we review the two specific scenarios 
of $A_\alpha$ discussed in Ref.~\cite{Fong:2010zu} and discuss an interesting point
missed by Ref.~\cite{Fong:2010zu} which actually allows for nonzero leading
$CP$ violation at zero temperature.

\section{The Lagrangian}
\label{sec:lag}

The superpotential for the type-I seesaw is given by 
\begin{eqnarray}
W_N & = & \frac{1}{2}M_i\hat{N_i^{c}}\hat{N_i^{c}}
+Y_{i\alpha}\hat{N_i^{c}}\hat{\ell}_{\alpha}\hat{H}_{u},
\end{eqnarray}
where $\hat{N_i^c}$, $\hat{\ell}_{\alpha}$ and $\hat{H}_{u}$ denote, respectively, 
the chiral superfields of the RHNs, the lepton doublet and the up-type Higgs doublet, and
$i$ and $\alpha$ are the RHN family and lepton flavor indices, respectively.
The $SU(2)_{L}$ contraction between $\hat{\ell}_{\alpha}$ 
and $\hat{H}_{u}$ is left implicit.
In the following, we will assume that the RHNs are hierarchical 
such that only the lightest RHN $N_1$ is relevant for soft leptogenesis.
Henceforth, we will drop the family index of RHN, for example,
$N \equiv N_1$ and $Y_{\alpha} \equiv Y_{1\alpha}$.
The corresponding soft terms are
\begin{eqnarray}
-{\cal L}_{{\rm soft}} & = & \widetilde{M}^{2}\widetilde{N}^{*}\tilde{N}
+\left(\frac{1}{2}BM\widetilde{N}\widetilde{N}
+A_\alpha \widetilde{N}\widetilde{\ell}_{\alpha}H_{u}+{\rm H.c.}\right).
\label{eq:soft}
\end{eqnarray}

The mass and interaction terms involving the
sneutrino $\widetilde{N}$ from $W_{N}$ are given by
\begin{eqnarray}
-{\cal L}_{\widetilde{N}} & = & \left|M\right|^{2}\widetilde{N}^{*}\tilde{N}
+\left(M^{*}Y_{\alpha}\widetilde{N}^{*}\widetilde{\ell}_{\alpha}H_{u}
+Y_{\alpha}\overline{\widetilde{H}_{u}^{c}}P_{L}\ell_{\alpha}\widetilde{N}+{\rm H.c.}\right),
\end{eqnarray}
where $P_{L,R}=\frac{1}{2}\left(1\mp\gamma_{5}\right)$. 
Through field redefinitions, it can be shown that the three
physical phases are
\be
\Phi_\alpha = \arg\left(A_\alpha Y_\alpha^* B^*\right).
\label{eq:phases}
\ee
Without loss of generality, the phases can be assigned to $A_\alpha$ 
and all other parameters will be taken real and positive.
We would like to stress that we do not assume the proportionality of $A_\alpha$
to the neutrino Yukawa couplings ($A_\alpha = A Y_\alpha$) as has been 
done in Refs. \cite{D'Ambrosio:2003wy,Grossman:2003jv,Fong:2008mu}
where there is only one physical phase $\Phi = \arg(AB^*)$.
As we will show in Sec. \ref{sec:CP_violation}, by considering generic $A_\alpha$ 
couplings, the $CP$ violation can be nonvanishing even at zero temperature.

Because of the bilinear $B$ term, $\widetilde{N}$ and $\widetilde{N}^{*}$ 
mix to form mass eigenstates
\begin{eqnarray}
\widetilde{N}_{+} & = & \frac{1}{\sqrt{2}}\left(\widetilde{N}+\widetilde{N}^{*}\right),\nonumber \\
\widetilde{N}_{-} & = & -\frac{i}{\sqrt{2}}\left(\widetilde{N}-\widetilde{N}^{*}\right),
\label{eq:mass_eigenstates}
\end{eqnarray}
with the corresponding masses given by 
\be
M_{\pm}^{2} =  M^{2}+\widetilde{M}^{2}\pm BM.
\label{eq:masses}
\ee
In order to avoid a tachyonic mass which implies an instability of 
the vacuum such that the sneutrino will develop a vacuum expectation value, 
we always assume $B < M + \widetilde M^2/M$. 

Rewriting the Lagrangian in terms of mass eigenstates $\widetilde{N}_{\pm}$
%Eq. (\ref{eq:mass_eigenstates}), 
we have
\begin{eqnarray}
-{\cal L}_{\widetilde{N}}-{\cal L}_{{\rm soft}} & = & 
M_{+}^{2}\widetilde{N}_{+}^{*}\widetilde{N}_{+}+M_{-}^{2}\widetilde{N}_{-}^{*}\widetilde{N}_{-}\nonumber \\
 &  & +\frac{1}{\sqrt{2}}\left\{ \widetilde{N}_{+}\left[ Y_{\alpha} \overline{\widetilde{H}_{u}^{c}}P_{L}\ell_{\alpha}
+\left(A_\alpha + M Y_\alpha\right)\widetilde{\ell}_{\alpha}H_{u}\right]\right.\nonumber \\
 &  & \left.+i \widetilde{N}_{-}\left[ Y_{\alpha} \overline{\widetilde{H}_{u}^{c}}P_{L}\ell_{\alpha}
+\left(A_\alpha - M Y_\alpha\right)\widetilde{\ell}_{\alpha}H_{u}\right]+{\rm H.c.}\right\} .
\label{eq:lag}
\end{eqnarray}

\subsection{General constraints}

The total decay width for $\widetilde{N}_\pm$ is given by
\be
\Gamma_{\pm} \simeq \frac{M}{4\pi}
\sum_\alpha \left[Y_\alpha^2 + \frac{|A_\alpha|^2}{2M^2} 
\pm \frac{Y_\alpha {\rm Re}(A_\alpha)}{M}
%- \frac{Y_\alpha B\, {\rm Re}(A_\alpha) }{2M^2}
\right],
\label{eq:decay_width}
%\mp \frac{|A_\alpha|^2 B}{4 M^3} \right],
\ee
where we have expanded up to ${\cal O}(Y_\alpha^2,m_{\rm SUSY}^2/M^2,Y_\alpha m_{\rm SUSY}/M)$ 
and ignored the final state phase space factors. 
We will impose the restriction that $|A_\alpha|, B < M$ and $Y_\alpha < 1$ 
to ensure that we are always in the perturbative regime. In principle, 
$m_{\rm SUSY}/M$ and $Y_\alpha$ can go up to $4\pi$ before perturbative theory 
breaks down but with our stronger restriction, we are not anywhere near 
the nonperturbative regime.

The out-of-equilibrium condition for leptogenesis is
\be
\Gamma_{\pm} \lesssim H(T = M),
\ee
where the Hubble expansion rate is given by 
$H = 1.66\sqrt{g_\star}\, T^2/M_{\rm Pl}$ with Planck mass 
$M_{\rm Pl} = 1.22 \times 10^{19}$ GeV. Assuming 
minimal supersymmetric Standard Model relativistic degrees of freedom,
we have $g_\star = 228.75$. The condition above translates to
\be
\sqrt{ \sum_\alpha\left[Y_\alpha^2 + \frac{|A_\alpha|^2}{2M^2} 
\pm \frac{Y_\alpha {\rm Re}(A_\alpha)}{M}\right] }
\lesssim 1.6 \times 10^{-5} \left(\frac{M}{10^7\,{\rm GeV}}\right)^{1/2}.
\label{eq:out-of-equilibrium}
\ee
From the condition above, we see that $|A_\alpha|$ is bounded
from above depending on $M$. For example if $M \sim $ TeV, we require
$|A_\alpha| \lesssim 10^{-4}$ GeV. At this low scale, the mass splitting 
between $\widetilde N_+$ and $\widetilde N_-$ is required to be of the order of
their decay widths such that the $CP$ violation is resonantly enhanced to 
yield successful leptogenesis \cite{Pilaftsis:1997jf,Pilaftsis:2003gt}. 
To avoid excessive fine-tuning, if we consider $|A_\alpha| \sim $ TeV, 
Eq.~\eqref{eq:out-of-equilibrium} implies $M \gtrsim 4\times 10^7\, {\rm GeV}$.

In type-I seesaw, barring special cancellation, we have the upper
bound on the sum of light neutrino masses from cosmology~\cite{Ade:2013zuv}
\begin{eqnarray}
\sum_\alpha \frac{Y_{\alpha}^{2}v_u^{2}}{M} & \lesssim & 
\sum_{i}m_{\nu_{i}}\simeq0.23\,{\rm eV},\nonumber \\
\sqrt{ \sum_\alpha Y_{\alpha}^2} & \lesssim & 
3\times10^{-4}\left(\frac{M}{10^{7}\,{\rm GeV}}\right)^{1/2}
\left(1 + \frac{1}{\tan^2\beta}\right)^{1/2},
\label{eq:Yboundcosmo}
\end{eqnarray}
where $\tan\beta\equiv v_u/v_d$ and $v_{u(d)} = \left<H_{u(d)}\right>$ are 
the up(down)-type Higgs vacuum expectation values.
$v_u^2 + v_d^2 = \sqrt{2}\,G_F^{-1} \simeq (174\,{\rm GeV})^2$ 
with $G_F$ the Fermi constant. For $\tan\beta \gtrsim 1$, the bound above is always 
less stringent than Eq.~\eqref{eq:out-of-equilibrium},
and, hence, the out-of-equilibrium condition alone is sufficient.

\section{$CP$ violation}
\label{sec:CP_violation}

In this section we will study $CP$ violation of the Lagrangian \eqref{eq:lag} 
from the interferences between tree-level and one-loop diagrams 
shown in Figs. \ref{fig:selfenergy_diagrams} and \ref{fig:vertex_diagrams}.
We will take into account thermal corrections while approximating 
sneutrinos $\widetilde N_\pm$ to always be at rest with respect 
to the thermal bath. Since we are in the regime where all three lepton flavors 
can be distinguished ($T \lesssim 10^9$ GeV), we will not sum over 
the lepton flavor in the final states \cite{Fong:2008mu}. 

To quantify the $CP$ violation, we define the $CP$ asymmetry 
for the decays $\widetilde N_\pm \to a_\alpha$ with
$a_\alpha = \{\widetilde\ell_\alpha H_u, \ell_\alpha \widetilde H_u\}$ as
\be
\epsilon^{S,V}_{\pm\alpha} \equiv 
\frac{\gamma(\widetilde N_\pm \to a_\alpha) 
- \gamma(\widetilde N_\pm \to \overline {a_\alpha})}
{\sum_{a_\beta;\beta}\left[\gamma(\widetilde N_\pm \to a_\beta) 
+ \gamma(\widetilde N_\pm \to \overline {a_\beta})\right]},
\label{eq:CP_asym}
\ee
where the superscripts $S$ and $V$ indicate the $CP$ violation 
coming from  self-energy and vertex corrections, 
respectively, $\overline {a_\alpha}$ indicates the $CP$ conjugate of $a_\alpha$,
and $\gamma(i \to j)$ is the thermal averaged
reaction density for the process $i \to j$ defined in Eq.~\eqref{eq:therm_ave_rate}. 
In the following, we will include the thermal effects associated 
with intermediate on-shell states which, as shown in Ref.~\cite{Garbrecht:2013iga}, 
will result in the cancellation of vertex $CP$ asymmetries from gaugino contributions 
\cite{Grossman:2004dz,Fong:2009iu}.
We will always approximate $\widetilde N_\pm$ to be at rest 
with respect to the thermal bath so that we can obtain analytical expressions for 
the $CP$ asymmetries (see Appendix \ref{app:therm}). 
Furthermore, we focus on the superequilibration regime which falls in the 
temperature range $T \lesssim 10^8 $ GeV for $m_{\rm SUSY} \sim 1$ TeV \cite{Fong:2010qh}.
The advantage is that in this regime, lepton and sleptons are not 
distinguished (they have the same chemical potentials) 
and so the two Boltzmann equations for the lepton asymmetry in particles  and 
sparticles can be reduced to one equation for the net lepton asymmetry.\footnote{We make 
the assumption of superequilibration also to highlight the positive effects of nonthermal $CP$ violation 
in soft leptogenesis. Including nonsuperequilibration effects, the efficiency of 
soft leptogensis is expected to be further enhanced, and this effect was studied 
in detail in Ref.~\cite{Fong:2010bv}.
The validity window of superequilibration can be enlarged by increasing the gaugino masses 
and $\mu$ parameter \cite{Fong:2010qh} and/or decreasing $|A_\alpha|$.}
Hence we are allowed to sum over $CP$ asymmetries of lepton and slepton final states as below.

\subsection{$CP$ violation from mixing}
\label{sec:CP_mixing}

\begin{figure}
\includegraphics[scale=0.6]{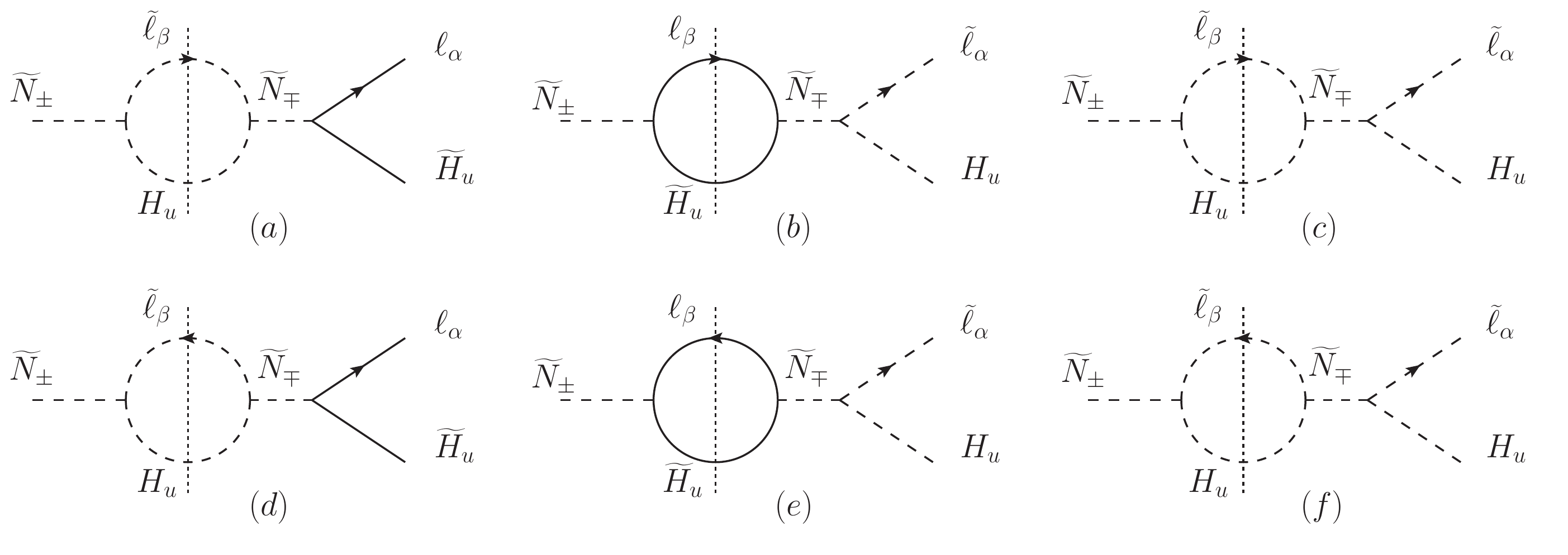}
\caption{One-loop self-energy diagrams for the decays 
$\widetilde N_\pm \to \ell_\alpha \widetilde H_u$ [(a),(d)] 
and $\widetilde N_\pm \to \widetilde\ell_\alpha H_u$ [(b),(c),(e),(f)].
The arrow indicates the flow of lepton number. 
The dotted vertical lines indicate the corresponding 
intermediate states go on mass shell.
The diagram with fermionic loop and fermionic 
final states does not contribute to the $CP$ violation since it does not 
involve the soft couplings $A_\alpha$.
\label{fig:selfenergy_diagrams}}
\end{figure}

In this subsection, we discuss the mixing $CP$ violation from self-energy corrections.
There are two kinds of self-energy diagrams as shown in Fig. \ref{fig:selfenergy_diagrams}: 
the diagrams with continuous flow of lepton number [Figs. 1(a)--1(c)]
and the diagrams with flow of lepton number inverted in the loop [Figs. 1(d)--1(f)]. 
Notice that diagrams with fermionic loop and fermionic final states do not contribute to 
the $CP$ violation since they do not involve the soft couplings $A_\alpha$.
From Figs. 1(a)--1(c), we obtain the respective 
contributions to the $CP$ asymmetries defined in Eq.~\eqref{eq:CP_asym} as follows\footnote{
The absorptive parts which regularize the singularity in the $\widetilde N_\pm$ propagators 
as $M_+ \to M_-$ are obtained by resumming self-energy diagrams following 
Refs.~\cite{Pilaftsis:1997jf,Pilaftsis:2003gt}.}:
\bea
\epsilon^{S,(a)}_{\pm\alpha} & = & \frac{1}{4\pi G_\pm(T)} 
Y_\alpha^2 \sum_\beta Y_\beta \frac{{\rm Im} (A_\beta)}{M}
\left(1 + \frac{\widetilde M^2}{M^2} \pm \frac{B}{M}\right)
\frac{2 B M}{4 B^2 + \Gamma_\mp^2} r_B(T) {c_F(T)} , \nonumber \\
\label{eq:ep_S_a} 
\epsilon^{S,(b)}_{\pm\alpha} & = & -\frac{1}{4\pi G_\pm(T)} 
Y^2  Y_\alpha \frac{{\rm Im} (A_\alpha)}{M} 
\left(1 + \frac{\widetilde M^2}{M^2} \pm \frac{B}{M}\right)
\frac{2 B M}{4 B^2 + \Gamma_\mp^2} r_F(T) c_B(T) ,  \\
\epsilon^{S,(c)}_{\pm\alpha} & = & \frac{1}{4\pi G_\pm(T)} \left[
\left(Y^2 - \sum_\beta \frac{|A_\beta|^2}{M^2} \right) 
Y_\alpha \frac{{\rm Im} (A_\alpha)}{M} 
- \left(Y_\alpha^2 - \frac{|A_\alpha|^2}{M^2} \right) 
\sum_\beta Y_\beta \frac{{\rm Im} (A_\beta)}{M} \right]\nonumber \\
& & \times \frac{2 B M}{4 B^2 + \Gamma_\mp^2} r_B(T) c_B(T) , \nonumber
\eea
where we define $Y^2 \equiv \sum_\alpha Y_\alpha^2$ and 
\be
\label{eq:G_T}
G_\pm(T) \equiv \left[Y^2 + \sum_\alpha \left( \frac{|A_\alpha|^2}{M^2} 
\pm \frac{2Y_\alpha {\rm Re}(A_\alpha)}{M} \right) \right] c_B(T)
+ Y^2 \left(1 + \frac{\widetilde M^2}{M^2} \pm \frac{B}{M} \right) c_F(T).
\ee
In the above $r_{B,F}(T)$ and $c_{B,F}(T)$ are temperature-dependent 
terms associated with intermediate on-shell and final states respectively, 
as given 
in Appendix \ref{app:therm}. We will also make use of the following identity
\be
r_F(T) c_B(T) = r_B(T) c_F(T),
\label{eq:therm_iden}
\ee
proven in Appendix \ref{app:therm}. Note that if we sum
over the lepton flavor $\alpha$ and use Eq.~\eqref{eq:therm_iden},
we obtain $\sum_\alpha \left( \epsilon^{S,(a)}_{\pm\alpha} 
+ \epsilon^{S,(b)}_{\pm\alpha} \right) 
= \sum_\alpha \epsilon^{S,(c)}_{\pm\alpha} = 0$, 
in agreement with the $T=0$ result of Ref.~\cite{Adhikari:2001yr} that if 
there is no $L$ violation to the right of the cut in the one-loop diagrams,
the net $CP$ violation on summing over all final states is zero.

From Figs. 1(d)--1(f), we have
\bea
\epsilon^{S,(d)}_{\pm\alpha} & = & \frac{1}{4\pi G_\pm(T)} 
Y_\alpha^2 \sum_\beta Y_\beta \frac{{\rm Im} (A_\beta)}{M}
\left(1 + \frac{\widetilde M^2}{M^2} \pm \frac{B}{M}\right)
\frac{2 B M}{4 B^2 + \Gamma_\mp^2} r_B(T) c_F(T) , \nonumber \\
\label{eq:ep_S_b}
\epsilon^{S,(e)}_{\pm\alpha} & = & \frac{1}{4\pi G_\pm(T)} 
Y^2  Y_\alpha \frac{{\rm Im} (A_\alpha)}{M} 
\left(1 + \frac{\widetilde M^2}{M^2} \pm \frac{B}{M}\right)
\frac{2 B M}{4 B^2 + \Gamma_\mp^2} r_F(T) c_B(T) , \\
\epsilon^{S,(f)}_{\pm\alpha} & = & \frac{1}{4\pi G_\pm(T)} \left[
- \left(Y^2 - \sum_\beta \frac{|A_\beta|^2}{M^2} \right) 
Y_\alpha \frac{{\rm Im} (A_\alpha)}{M} 
- \left(Y_\alpha^2 - \frac{|A_\alpha|^2}{M^2} \right) 
\sum_\beta Y_\beta \frac{{\rm Im} (A_\beta)}{M} \right]\nonumber \\
& & \times \frac{2 B M}{4 B^2 + \Gamma_\mp^2} r_B(T) c_B(T) . \nonumber
\eea
Notice the leading contributions from $\widetilde N_+$ 
and $\widetilde N_-$ in Eqs.~\eqref{eq:ep_S_a} and \eqref{eq:ep_S_b} 
come with the same sign and, hence, they 
will contribute constructively to the lepton number asymmetry.

The total $CP$ asymmetry from mixing $\epsilon^S_{\pm \alpha} \equiv 
\sum_{n=\{ a,b,c,d,e,f \}}\epsilon^{S,(n)}_{\pm \alpha}$ 
is given by
\bea
\epsilon^S_{\pm \alpha}
& = & \frac{1}{4\pi G_\pm (T)} 
Y_\alpha^2 \sum_\beta Y_\beta \frac{{\rm Im} (A_\beta)}{M}
\frac{4 B M}{4 B^2 + \Gamma_\mp^2} \left[c_F(T) - c_B(T)\right] r_B(T) \nonumber \\
\label{eq:ep_S_tot}
& & + \,\frac{1}{4\pi G_\pm(T)} 
\frac{|A_\alpha|^2}{M^2} \sum_\beta Y_\beta \frac{{\rm Im} (A_\beta)}{M}
\frac{4 B M}{4 B^2 + \Gamma_\mp^2} r_B(T) c_B(T)  \\
& & + \,\frac{1}{4\pi G_\pm(T)} 
Y_\alpha^2 \sum_\beta Y_\beta \frac{{\rm Im} (A_\beta)}{M}
\left(\frac{\widetilde M^2}{M^2} \pm \frac{B}{M}\right)
\frac{4 B M}{4 B^2 + \Gamma_\mp^2} r_B(T) c_F(T). \nonumber
\eea 
In the above, the first term vanishes in the zero temperature limit
$T \to 0$ when $c_{B,F}(T) \to 1$ and $r_{B,F}(T) \to 1$, 
while the terms higher order in $m_{\rm SUSY}/M$ survive. 
They remain nonzero after summing over the lepton flavor $\a$.
In the following in order to make the dependence of thermal and nonthermal $CP$ asymmetries 
in Eq.~\eqref{eq:ep_S_tot} on the model parameters more transparent, 
it is instructive to look at two limiting cases (i) $Y_\alpha \gg A_\alpha/M$
and (ii) $Y_\alpha \ll A_\alpha/M$ where in case (i), the thermal $CP$
violation dominates, while in case (ii), the nonthermal $CP$ violation dominates.\footnote{By 
thermal (nonthermal) $CP$ violation, we refer to the case 
where $CP$ violation does (not) vanish as $T \to 0$.}

\begin{itemize}

\item
In the limit (i) $Y_\alpha \gg A_\alpha/M$, we have
\bea
\epsilon^S_{\pm \alpha}
& \simeq & \frac{1}{4\pi}
P_\alpha \sum_\beta Y_\beta \frac{{\rm Im} (A_\beta)}{M}
\frac{4 B M}{4 B^2 + \Gamma_Y^2} \frac{c_F(T) - c_B(T)}{c_F(T) + c_B(T)} r_B(T),
\label{eq:ep_S_tot_i}
\eea
where we define the flavor projector 
$P_\alpha \equiv Y_\alpha^2/Y^2$ with $\sum_\alpha P_\alpha = 1$
and $\Gamma_Y \equiv \frac{Y^2 M}{4 \pi}$,
and we have dropped the terms higher order in $m_{\rm SUSY}/M$.
In this case, the $CP$ asymmetry in Eq.~\eqref{eq:ep_S_tot_i} 
is proportional to $c_F(T) - c_B(T)$ which goes to zero as $T \to 0$, and, 
hence, the contribution to the $CP$ violation is the thermal one.

In the resonant regime where $B \sim \Gamma_\pm$, we have 
$\epsilon^S_{\pm} \sim (|A|/M)/Y$ where we have suppressed the
lepton flavor index for an order of magnitude estimation.
In this case, a large $\epsilon^S_\pm$ can be
obtained which allows TeV-scale leptogenesis but at the cost of 
having unnaturally small $|A|, B \ll$ TeV. 

Away from the resonant regime when $B \gg \Gamma_\pm$, 
the $CP$ asymmetries go as $\epsilon^S_{\pm} \sim 10^{-1}Y |A|/B$ 
assuming ${\cal O}(1)$ contribution from the $CP$ phases of Eq.~\eqref{eq:phases}. 
Taking $|A| \sim \,{\rm TeV} \gtrsim B$
together with the out-of-equilibrium decay condition \eqref{eq:out-of-equilibrium} 
gives us sufficient $CP$ asymmetries $\epsilon^S_{\pm} \gtrsim 10^{-6}$
for $M \gtrsim 10^7$ GeV.

\item 
In the other limit (ii) $Y_\alpha \ll A_\alpha/M$, 
we have
\bea
\epsilon^S_{\pm \alpha} 
& \simeq & \frac{1}{4\pi}\frac{|A_\alpha|^2}{\sum_\delta |A_\delta|^2}  
\sum_\beta Y_\beta \frac{{\rm Im} (A_\beta)}{M} 
\frac{4 B M}{4 B^2 + \Gamma_A^2} r_B(T),
\label{eq:ep_S_tot_ii}
\eea
where $\Gamma_A \equiv \sum_\alpha \frac{|A_\alpha|^2}{8\pi M}$.
The $CP$ asymmetries Eq.~\eqref{eq:ep_S_tot_ii} clearly do not vanish 
at $T = 0$, and this represents a \emph{nonthermal} $CP$ violation.
Of course thermal effects are always there but the fact that the CP violation 
is nonvanishing at $T = 0$ implies that it is less suppressed compared 
to case (i).

In the resonant regime $B \sim \Gamma_\pm$, we have 
$\epsilon^S_{\pm} \sim Y/(|A|/M)$.
In this case too a large $\epsilon^S_\pm$ can be
obtained which allows TeV-scale leptogenesis but at the cost of 
having unnaturally small $|A|, B \ll$ TeV. 

Away from the resonant regime with $B \gg \Gamma_\pm$, 
the $CP$ asymmetries, like in the limit (i), 
go as $\epsilon^S_{\pm} \sim 10^{-1}Y |A|/B$ assuming 
${\cal O}(1)$ contribution from the CP phases of Eq.~\eqref{eq:phases}.
%order of one CP phases in Eq.~\eqref{eq:phases}.
Hence taking $|A| \sim \,{\rm TeV} \gtrsim B$
together with the out-of-equilibrium decay condition \eqref{eq:out-of-equilibrium} 
gives us sufficient $CP$ asymmetries $\epsilon^S_{\pm} \gtrsim 10^{-6}$
for $M \gtrsim 10^7$ GeV.

\end{itemize}

\begin{figure}
\includegraphics[scale=0.4]{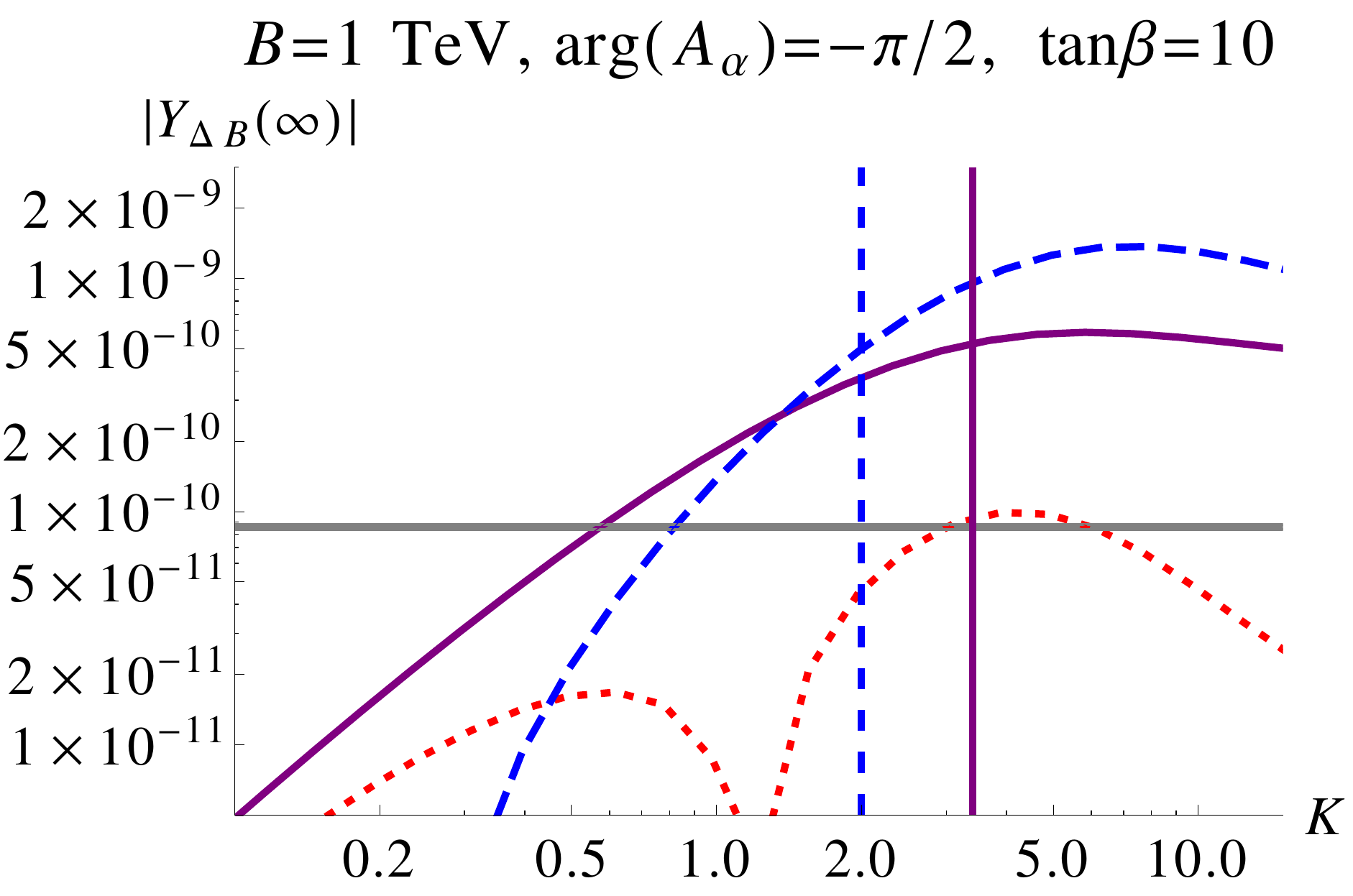}
\includegraphics[scale=0.4]{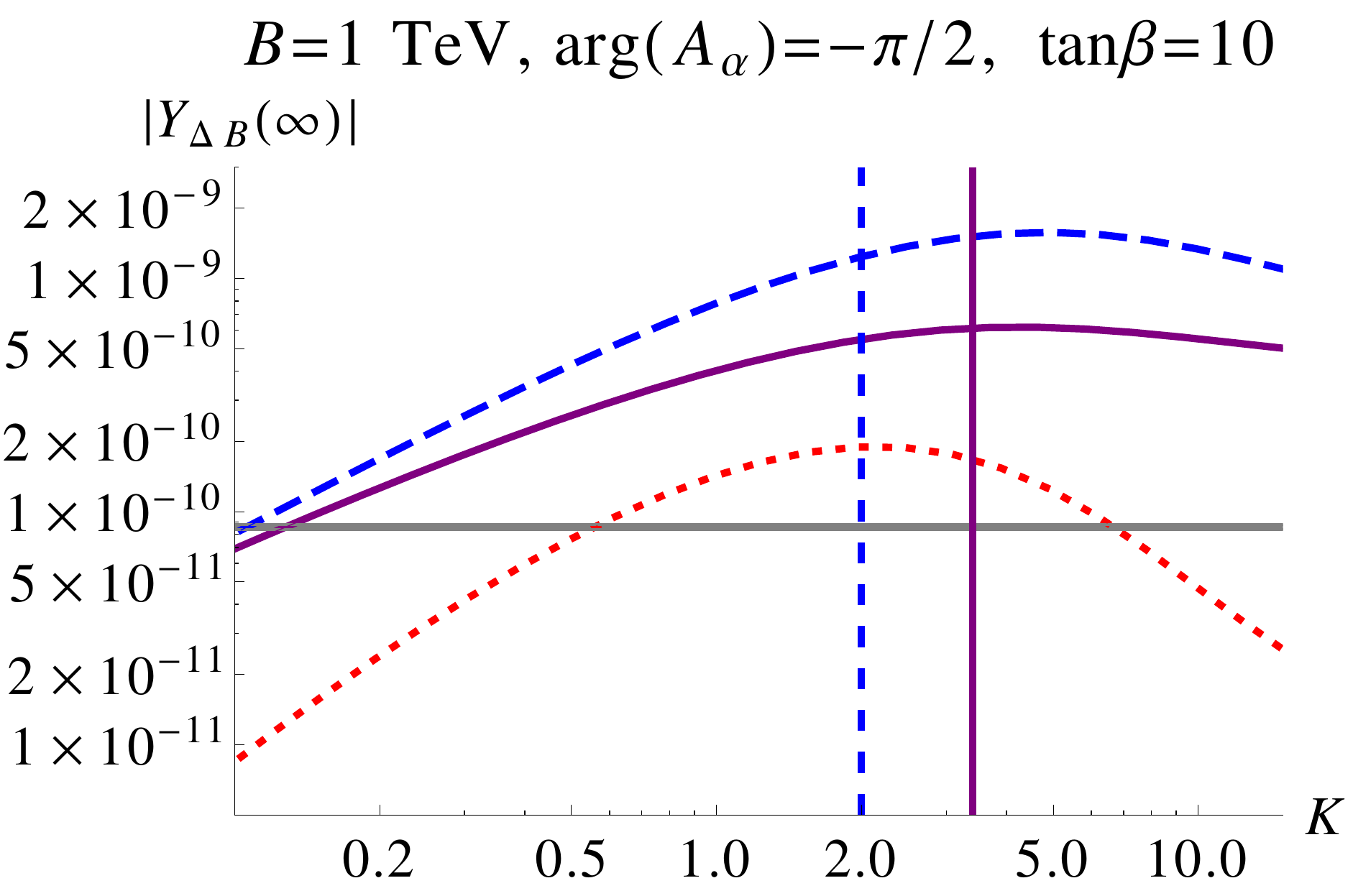}
\caption{
The absolute value of the final baryon asymmetry $|Y_{\Delta B}(\infty)|$
as a function of the washout parameter $K \equiv \Gamma_\pm/H(T=M)$ for
$M = 5 \times 10^7$ GeV for the three scenarios described in the
text: NTD (blue dashed), TD (red dotted) and MIX (purple solid).
The left plot corresponds to the case of zero initial number densities 
of $N$ and $\widetilde N_\pm$, while the right plot corresponds 
to the case of thermal initial number densities of $N$ and $\widetilde N_\pm$.
The regions to the left of the blue dashed and purple solid 
vertical lines correspond to $K$ values when $A_\alpha < 5$ TeV 
for the NTD and MIX scenarios, respectively, while for the TD scenario
we always have $A_\alpha < 5$ TeV in the range of the plot.
The gray band represents the recent combined Planck and WMAP CMB
 measurements of cosmic baryon asymmetry~\cite{Ade:2013zuv,Bennett:2012zja} at 2$\sigma$.
 The dip in the TD scenario in the left plot refers to a change in the sign of the baryon asymmetry.
\label{fig:K_YB}}
\end{figure}

\begin{figure}
\includegraphics[scale=0.7]{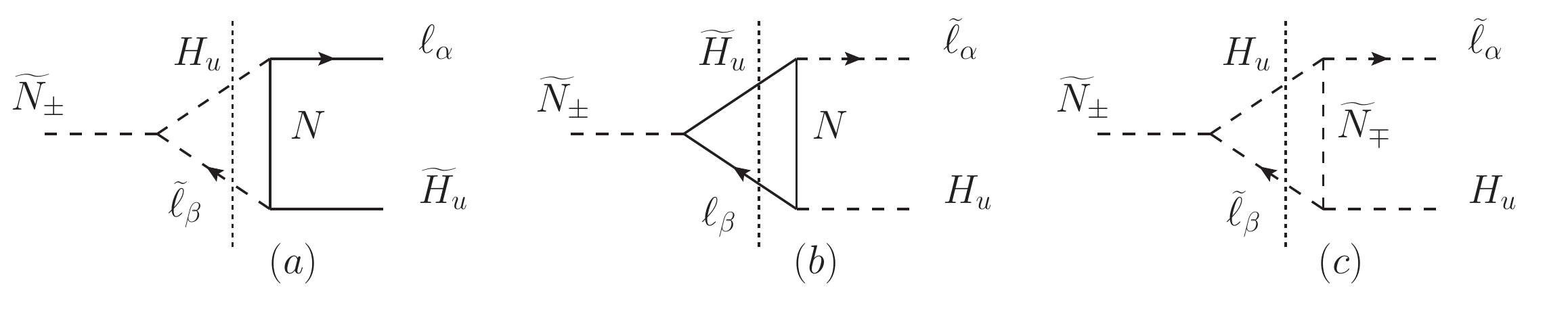}
\caption{
One-loop vertex diagrams for the decays 
$\widetilde N_\pm \to \ell_\alpha \widetilde H_u$ [(a)]
and $\widetilde N_\pm \to \widetilde\ell_\alpha H_u$ [(b),(c)]
with the conventions of Fig. \ref{fig:selfenergy_diagrams}.
The diagrams with fermionic loop and fermionic 
final states do not contribute to the $CP$ violation since they do not 
involve the soft couplings $A_\alpha$. 
\label{fig:vertex_diagrams}}
\end{figure}

To confirm our estimation of successful leptogenesis and also 
to illustrate the enhancing effects of nonthermal $CP$ violation, we numerically
solve the Boltzmann equations using the expression for the asymmetry parameter
in Eq.~\eqref{eq:ep_S_tot}. For simplicity, we consider only
decays and inverse decays of $N$ and $\widetilde N_\pm$. 
We will also define the washout parameter as $K \equiv \Gamma_\pm/H(T=M)$
with $\Gamma_\pm$ given by Eq. \eqref{eq:decay_width}. 
In Fig. \ref{fig:K_YB}, we plot the absolute value of the final baryon asymmetry
$|Y_{\Delta B}(\infty)|$ as a function of $K$ for the following three scenarios: 

\begin{itemize}

\item Non Thermal dominated (NTD): 
In this scenario, we choose {\boldmath$A$}$/M=(10^{-4},10^{-2},1) w$ 
and {\boldmath$Y$}$=(10^{-5},10^{-3},10^{-1}) w$.

\item Thermal Dominated (TD):
In this scenario, we choose {\boldmath$A$}$/M=(10^{-5},10^{-3},10^{-1}) w$ 
and {\boldmath$Y$}$=(10^{-4},10^{-2},1) w$.

\item Mixed (MIX): 
In this scenario, we choose {\boldmath$A$}$/M=(10^{-4},10^{-2},1) w$ 
and {\boldmath$Y$}$=(10^{-4},10^{-2},1) w$.

\end{itemize}
In the above {\boldmath$A$} and {\boldmath$Y$} represent the couplings written as 3-vectors.
In all the scenarios above, we vary $w$ between $10^{-6}$ and $10^{-4}$ 
such that we scan through the parameter space from the weak washout ($K = 0.1$) 
to the strong washout ($K = 15$) regime while still respecting 
the cosmological bound on the sum of light neutrino masses in Eq. \eqref{eq:Yboundcosmo}.
For definiteness, we also fix $M = 5 \times 10^7$ GeV, $\tan \beta = 10$, 
$\arg(A_\alpha) = -\pi/2$, and $B = 1$ TeV. 

In Fig. \ref{fig:K_YB}, for the left plot, 
we solve from an initial time with $T\gg M$ assuming zero initial number densities for 
$N$ and $\widetilde N_\pm$ while for the right plot, we
assume thermal initial number densities for $N$ and $\widetilde N_\pm$.
%To confront 
For the observed baryon asymmetry, we use the recent combined 
Planck and WMAP CMB measurements of cosmic baryon asymmetry
~\cite{Ade:2013zuv,Bennett:2012zja} at 2$\sigma$,
\begin{equation}
  \label{eq:CMB}
  Y_{\Delta B}^{\rm CMB}=(8.58\pm 0.22)\times 10^{-11}\,,
\end{equation}
which is plotted as the gray band in Fig. \ref{fig:K_YB}.
From the plots, we see that in the NTD (blue dashed line)
and MIX (purple solid line) scenarios, $|Y_{\Delta B}(\infty)|$ falls off
very slowly in the strong washout regime $K > 1$. 
The reason is that the falloff in their efficiencies
is almost completely compensated by the increases in their respective $CP$ asymmetries
as one increases $A_\alpha/M$. On the other hand, in the TD (red dotted line) scenario,
in the $K > 1$ regime, $|Y_{\Delta B}(\infty)|$ falls off much faster due to
the additional suppression from the partial cancellation between
the $CP$ asymmetries from the decays of $\widetilde N_\pm$ to scalars and fermions.

In our study we are also interested in the situation 
when $A_\alpha$ and $B$ are restricted to be around the TeV scale.
In Fig. \ref{fig:K_YB}, the regions to the left of the thick blue dashed and 
purple solid vertical lines correspond to $K$ when $A_\alpha < 5$
TeV for the NTD and MIX scenarios, respectively, while 
$A_\alpha < 5$ TeV for the TD scenario in the entire range of $K$ considered in the plot.
We find, for example, for the case of zero initial number densities of $N$ and $\widetilde N_\pm$, 
the correct amount of baryon asymmetry can be obtained
for NTD at $K \sim 0.8$, MIX at $K \sim 0.6$, and TD at $K \sim 4$.
Notice that the appropriate sign baryon asymmetry can always be
obtained by choosing the appropriate phases of the complex couplings $A_\alpha$.
From this numerical exercise we conclude that generation of sufficient baryon asymmetry
is possible for TeV-scale $A_\alpha$ and $B \gg \Gamma_\pm$, i.e., far away from the resonant regime.
Besides, we also see that nonthermal $CP$ violation can significantly
enhance the efficiency of soft leptogenesis.

Finally, in Appendix \ref{app:CP_asym} we will discuss two special cases, namely,
(a) $A_\alpha = A Y_\alpha$ and (b) $A_\alpha = A Y^2/(3Y_\alpha)$
considered in previous work.

\subsection{$CP$ violation from vertex corrections}
\label{sec:CP_vertex}

In this subsection, we discuss the $CP$ violation from vertex corrections.
From Figs. 3(a)--3(c), we obtain 
\bea
\epsilon^{V,(a)}_{\pm\alpha} & = & \mp\frac{1}{8\pi G_\pm(T)} 
Y_\alpha^2 \sum_\beta Y_\beta \frac{{\rm Im} (A_\beta)}{M}
\ln\frac{M_\pm^2 + M^2}{M^2} r_B(T) c_F(T) , \nonumber \\
\label{eq:ep_V}
\epsilon^{V,(b)}_{\pm\alpha} & = & \mp\frac{1}{8\pi G_\pm(T)} 
Y^2 Y_\alpha \frac{{\rm Im} (A_\alpha)}{M} 
\ln\frac{M_\pm^2 + M^2}{M^2} r_F(T) c_B(T) , \\
\epsilon^{V,(c)}_{\pm\alpha} & = & \pm\frac{1}{8\pi G_\pm(T)} \left[
\left(Y^2 - \sum_\beta \frac{|A_\beta|^2}{M^2} \right) 
Y_\alpha \frac{{\rm Im} (A_\alpha)}{M} 
+ \left(Y_\alpha^2 - \frac{|A_\alpha|^2}{M^2} \right) 
\sum_\beta Y_\beta \frac{{\rm Im} (A_\beta)}{M} \right]\nonumber \\
& & \times \frac{M^2}{M_\pm^2}\ln\frac{M_\pm^2 + M_\mp^2}{M_\mp^2} r_B(T) c_B(T).\nonumber
\eea
Summing over the contributions above and expanding in $B/M \ll 1$
in the numerators, we have
\bea
\epsilon^{V}_{\pm \alpha} & \equiv & 
\epsilon^{V,(a)}_{\pm \alpha} + \epsilon^{V,(b)}_{\pm \alpha} 
+ \epsilon^{V,(c)}_{\pm \alpha} \nonumber \\
\label{eq:ep_V_tot}
& = & \mp \frac{\ln 2}{8\pi G_\pm(T)} \left[
Y^2 Y_\alpha \frac{{\rm Im} (A_\alpha)}{M} 
+ Y_\alpha^2 \sum_\beta Y_\beta \frac{{\rm Im} (A_\beta)}{M} \right]
\left[c_F(T) - c_B(T) \right] r_B(T)  \nonumber \\
& & -\frac{1}{8\pi G_\pm(T)} \left[
Y^2 Y_\alpha \frac{{\rm Im} (A_\alpha)}{M} 
+ Y_\alpha^2 \sum_\beta Y_\beta \frac{{\rm Im} (A_\beta)}{M} \right]
\frac{B}{M} \left[\frac{c_F(T)}{2} + (\ln 2 -1 ) c_B(T) \right] r_B(T)\\
& & \mp \, \frac{\ln 2}{8\pi G_\pm(T)} \left[
 \sum_\beta \frac{|A_\beta|^2}{M^2}Y_\alpha \frac{{\rm Im} (A_\alpha)}{M} 
+ \frac{|A_\alpha|^2}{M^2} \sum_\beta Y_\beta \frac{{\rm Im} (A_\beta)}{M} \right]
 r_B(T) c_B(T) \nonumber \\
& & +\, \frac{1}{8\pi G_\pm(T)} \left[
 \sum_\beta \frac{|A_\beta|^2}{M^2}Y_\alpha \frac{{\rm Im} (A_\alpha)}{M} 
+ \frac{|A_\alpha|^2}{M^2} \sum_\beta Y_\beta \frac{{\rm Im} (A_\beta)}{M} \right]
 \frac{B}{M} (\ln 2 -1) r_B(T) c_B(T).\nonumber
\eea
The leading contributions from $\widetilde N_\pm$ [first and third lines of 
Eq. \eqref{eq:ep_V_tot}] come at the order of $\epsilon^V\sim 10^{-1}Y^2$ 
[taking $Y_\alpha\sim{\rm Im}(A_\alpha)/M$], which are too small for successful leptogenesis 
from Eq. (\ref{eq:out-of-equilibrium}) for $M\gsim 10^7$ GeV. 
Of course the same conclusion holds also when $Y_\alpha\gg{\rm Im}(A_\alpha)/M$
or $Y_\alpha\ll{\rm Im}(A_\alpha)/M$. 
Besides, notice also that the leading contributions from $\widetilde N_\pm$ 
come with the opposite signs, and, hence, they will contribute destructively 
to the total lepton number asymmetry. Upon expanding $G_\pm(T)$ terms also in $B/M \ll 1$, 
we obtain an additional suppression factor $B/M$ like the terms in the 
second and fourth lines in Eq.~\eqref{eq:ep_V_tot}.
Hence we conclude that the vertex CP violation is irrelevant 
for soft leptogenesis.

\begin{figure}
\includegraphics[scale=0.7]{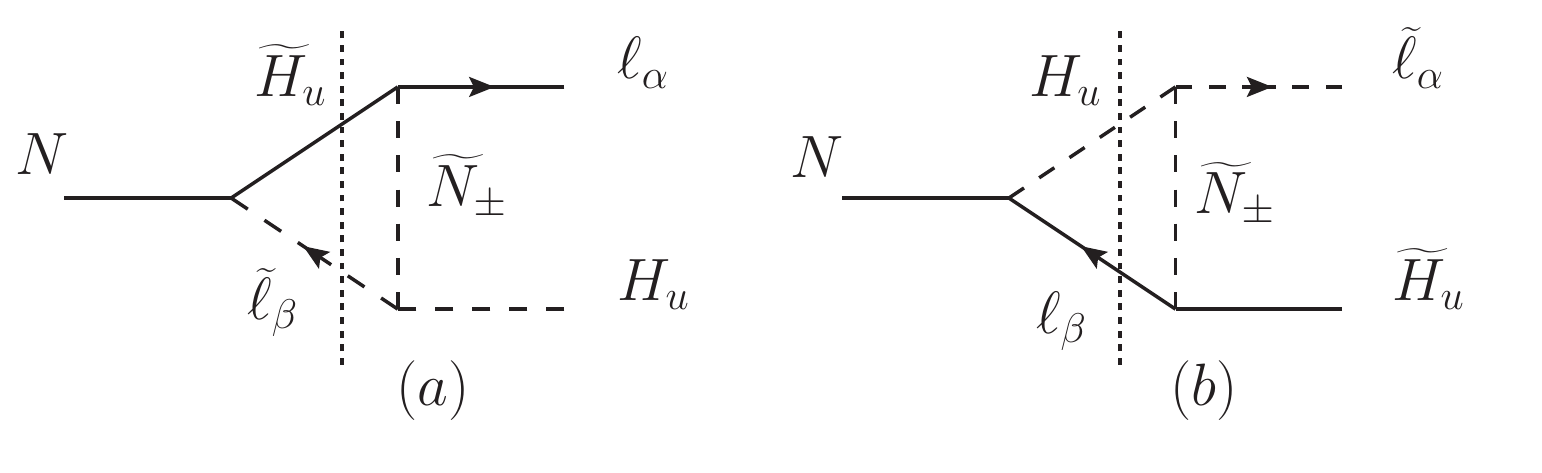}
\caption{One-loop vertex diagrams for the decays 
$N \to \ell_\alpha H_u$ [(a)] 
and $N \to \widetilde\ell_\alpha \widetilde H_u$ [(b)]
with the conventions of Fig. \ref{fig:selfenergy_diagrams}.
\label{fig:N_vertex_diagrams}}
\end{figure}

So far, we have been discussing the contributions 
of soft terms to $CP$ violation in the decay of $\widetilde N_\pm$.
In fact the soft terms also provide new sources of $CP$ violation 
in the one-loop vertex diagrams for the decays of the heavy neutrino $N$ 
as shown in Fig. \ref{fig:N_vertex_diagrams}.
Nevertheless the $CP$ violation from these diagrams comes at the same order as 
Eq.~\eqref{eq:ep_V_tot} and, hence, is too small for successful leptogenesis.

\section{Phenomenological constraints}
\label{sec:pheno}

We are primarily concerned with scenarios with $M\gsim 10^7\gev$
for which the production of sneutrinos is beyond the energy range of current colliders.  
However, even if $M_\pm \sim$ TeV, the bound on the Yukawa couplings from the requirement 
of out-of-equilibrium decays of $\widetilde N_\pm$ [in Eq. \eqref{eq:out-of-equilibrium}] 
makes $\widetilde N_\pm$ impossible to be produced at colliders \cite{Bambhaniya:2014hla}.
On the other hand, the soft SUSY breaking parameters 
relevant for soft leptogenesis $A_\alpha$, $B$, and $\widetilde M$ 
can contribute to Electric Dipole Moments (EDM) of leptons and 
to Charged Lepton Flavor Violating (CLFV) interactions though
the analysis of Ref.~\cite{Kashti:2004vj} under the assumption of 
universality soft trilinear couplings $A_\alpha = A Y_\alpha$ 
showed that the contributions to EDM and CLFV are much below 
the experimental bounds. Here we will repeat the analysis of 
Ref.~\cite{Kashti:2004vj} considering a generic $A_\alpha$.
In Ref.~\cite{Dedes:2007ef}, the phenomenological consequences 
of the soft terms considering three generations of RHN chiral 
superfields have been discussed at length. 
Clearly, these soft parameters are connected with the mechanism of 
SUSY breaking and as such are model dependent. 
Here we will remain agnostic about the SUSY breaking mechanism and simply 
focus on the phenomenological constraints on these parameters and, in particular,
we will focus only on parameters related to $N_1$ which are relevant for 
soft leptogenesis, i.e., $B$, $\widetilde M$, $A_\alpha$, $Y_\alpha$, and $M$.
Without fine-tuning, we consider the soft parameters $B$, $\widetilde M$, and 
$A_\alpha$ to be similar or smaller than $m_{\rm SUSY} \sim$ TeV. 
On the other hand, the parameters $Y_\alpha$ and $M$ are subject only to the
out-of-equilibrium $\widetilde N_\pm$ decay constraint in
Eq.~\eqref{eq:out-of-equilibrium} 
and less stringently to the cosmological bound on the sum of neutrino 
masses in Eq.~\eqref{eq:Yboundcosmo}. The running of $Y_\alpha$ from the high scale 
down to the weak scale gives some corrections at the level of $10\%-20\%$ 
(see Fig. 3 of Ref. \cite{Giudice:2003jh})
which we will ignore in the following.

\subsubsection{Electric dipole moment of the electron}
Assuming ${\cal O}(1)$ contribution of the
phases and mixing angles in the chargino sectors, 
the contributions of $A_\alpha$ and $B$ to the EDM of the electron 
are given by \cite{Kashti:2004vj}
\be
|d_e|\approx \frac{e\,m_e \tan\beta}{16 \pi m^2_{\tilde\nu}}
\left|\frac{m_\chi Y_\alpha}{M^2}\right|
\left(
|A_\alpha| + {B Y_\alpha} \right)\,,
\label{eq:EDM}
\ee
where $m_e$ is the electron mass, $m_{\tilde\nu}^2$ is the squared mass 
of the light sneutrino and $m_\chi$ the mass of chargino. 
For generic $A_\alpha$, the first term in in Eq.~\eqref{eq:EDM} dominates. 
Taking $m_{\tilde\nu} = m_\chi = m_{\rm SUSY}$ and making use of
Eq.~\eqref{eq:out-of-equilibrium}, we have
\be
|d_e| \lesssim 5 \times 10^{-38} \left(\frac{\tan\beta}{10}\right)
\left(\frac{10^7\,{\rm GeV}}{M}\right)^{3/2} 
\left(\frac{1\,{\rm TeV}}{m_{\rm SUSY}}\right) e\,{\rm cm},
\ee
which is much stronger than the current experimental bound 
$|d_e|_{\rm exp} < 8.7 \times 10^{-29}e\,\rm{cm}$ \cite{Baron:2013eja}. 
The contributions to $\mu$ and $\tau$ EDM can be estimated
by replacing $m_e$ in Eq.~\eqref{eq:EDM} by $m_\mu$ 
and $m_\tau$, respectively, but the current experimental constraints
on them are still a lot weaker:
$|d_\mu|_{\rm exp} < 1.9 \times 10^{-19}e\,\rm{cm}$ \cite{Bennett:2008dy} 
and $|d_\tau|_{\rm exp} < 5.1 \times 10^{-17}e\,\rm{cm}$ \cite{Inami:2002ah}.

\subsubsection{Charged lepton flavor violating interactions}

The branching ratio for charged lepton flavor violations 
due to nonvanishing off-diagonal elements of the soft mass matrix 
of the doublet sleptons $m_{\tilde \ell}^2$ is given by \cite{Kashti:2004vj,Hirsch:2012ti}
\be
\mbox{BR}(\ell_\alpha \rightarrow \ell_\beta \gamma) \approx \frac{\alpha^3}{G_F^2}
\frac{ \left| (m_{\tilde \ell}^2)_{\alpha\beta} \right|^2  }{m_{\rm SUSY}^8} \tan^2\beta,
\label{eq:CLFV}
\ee
where $\alpha$ is the fine structure constant.
In general, the off-diagonal elements of $m_{\tilde \ell}^2$
will induce too-large CLFV rates. The usual solution is to assume mSUGRA
boundary conditions at the grand unified theory (GUT) scale where 
the off-diagonal elements of $m_{\tilde \ell}^2$ vanish. 
In this case, as $m_{\tilde \ell}^2$ evolves from the GUT scale $M_{\rm GUT}$
to the RHN mass scale $M$, the off-diagonal elements will be generated
due to the renormalization effects as \cite{Hisano:1995cp}
\be
(m_{\tilde \ell}^2)_{\alpha\beta} 
\approx -\frac{1}{8\pi^2} A_\alpha^* A_\beta 
\ln \left(\frac{M_{\rm GUT}}{M}\right)
\label{eq:od_slepton_masses}
\ee
for $\alpha \neq \beta$, and we have kept only the dominant contributions
from $A_\alpha$.

The most stringent constraint on the rare decay $\mu \rightarrow e \gamma$ 
comes from the nonobservation of the process from the 
MEG experiment \cite{Adam:2011ch,Adam:2013mnn} 
which has set the new bound on the branching ratio for $\mu \to e \gamma$,
\be
\mbox{BR}(\mu \rightarrow e \gamma)_{\rm exp} < 5.7 \times 10^{-13}.
\label{eq:mu_to_egamma}
\ee
Substituting Eq.~\eqref{eq:od_slepton_masses} in 
Eq.~\eqref{eq:CLFV} and applying the constraint Eq.~\eqref{eq:mu_to_egamma},
we obtain
\be
|A_\mu^* A_e| \lesssim 5 \times 10^3\,{\rm GeV}^2
\left(\frac{m_{\rm SUSY}}{1\,{\rm TeV}}\right)^4
\left(\frac{10}{\tan\beta}\right),
\label{eq:CLFV_mu}
\ee
where we have taken $M_{\rm GUT} = 10^{16}$ GeV and $M = 10^7$ GeV.
Similarly using the experimental bounds on CLFV in $\tau$ decays,
$\mbox{BR}(\tau \rightarrow e \gamma)_{\rm exp} < 3.3 \times 10^{-8}$ and
$\mbox{BR}(\tau \rightarrow \mu \gamma)_{\rm exp} < 4.4 \times 10^{-8}$ \cite{Aubert:2009ag},
we obtain
\be
|A_\tau^* A_e| \approx |A_\tau^* A_\mu| \lesssim 1 \times 10^6\,{\rm GeV}^2 
\left(\frac{m_{\rm SUSY}}{1\,{\rm TeV}}\right)^4
\left(\frac{10}{\tan\beta}\right).
\label{eq:CLFV_tau}
\ee
For $m_{\rm SUSY}$ at the TeV scale, the bound \eqref{eq:CLFV_tau} can be satisfied
with $A_\alpha$ also at the TeV scale while the stronger bound \eqref{eq:CLFV_mu} 
requires either $A_e$ and/or $A_\mu$ to be smaller than TeV scale. 
As discussed in Sec. \ref{sec:CP_mixing}, 
the mixing $CP$ asymmetries away from the resonant regime
go as $\epsilon^S_{\alpha} \sim 10^{-1}Y_\alpha |A_\alpha|/B$
and can be large enough with $M \gtrsim 10^7$ GeV
and having one of the $A_\alpha \sim {\rm TeV} \gtrsim B$.

In addition, the off-diagonal entries of the slepton mass matrix 
can also give rise to other CLFV processes like 
$\mu \rightarrow 3 e$ and $\mu - e $ conversion. 
If such processes are dominated by the dipole-type operator 
for relatively large $\tan \beta$, $\mbox{BR}(\mu \rightarrow 3 e)$ 
and the rate of  $\mu - e $ conversion rate $R_{\mu e}$ are proportional to 
$\mbox{BR}(\mu \rightarrow e \gamma)$ and are approximately given by \cite{Ellis:2002fe}
\begin{eqnarray}
\mbox{BR}(\mu \rightarrow 3 e) \sim 6.6 \times 10^{-3} \mbox{BR}(\mu \rightarrow e \gamma),
\end{eqnarray}
and, for the $^{27}_{13}$Al nucleus, by \cite{Kitano:2002mt}
\begin{eqnarray}
R_{\mu e} \sim 2.5 \times 10^{-3} \mbox{BR}(\mu \rightarrow e \gamma).
\end{eqnarray}
The present constraints coming from these CLFV processes are 
less severe than those coming from $\mu \rightarrow e \gamma $. 
However, in future experiments the sensitivity for such processes may 
improve, which could constrain the presently allowed parameter space or lead to a detection of 
such lepton flavor violating processes. As for example,
the future Mu3e experiment \cite{Blondel:2013ia} could reach a sensitivity 
of $\sim 10^{-15} -- 10^{-16}$ for $\mbox{BR}(\mu \rightarrow 3 e)$. For the $\mu -e$
conversion process, from the Mu2e \cite{Abrams:2012er} and COMET  \cite{Kuno:2012pt} 
experiments, the bound could reach the level of $R_{\mu e} \sim 10^{-17} $ for 
the $^{27}_{13}$Al nucleus, while the PRISM/PRIME \cite{Kuno:2012pt} project 
may have 2 orders of magnitude greater sensitivity.

\section{Conclusions}
\label{sec:conclusions}

In the framework of local SUSY, soft leptogenesis is an attractive 
mechanism to explain the cosmological matter-antimatter asymmetry since
it works at the lower temperature regime $T \lesssim 10^9$ GeV 
where the conflict with the overproduction of gravitinos can be relaxed 
or even evaded. We showed that by considering generic soft trilinear $A_\alpha$ couplings 
there are two interesting consequences:
(1) one can realize nonthermal $CP$ violation where the CP asymmetries in 
the decays of heavy sneutrinos to lepton and sleptons do not cancel 
at zero temperature resulting in an enhanced efficiency in generating baryon asymmetry, and
(2) the dominant $CP$ violation from self-energy corrections is sufficient 
even far away from the resonant regime and the relevant soft parameters
can assume \emph{natural} values at around the TeV scale.
For successful soft leptogenesis, we considered two requirements: 
the out-of-equilibrium decays of heavy sneutrinos and a large 
enough $CP$ violation. Assuming $m_{\rm SUSY} \sim$ TeV, we found the 
following conditions $A_\alpha \sim \,{\rm TeV} \gtrsim B$ and $M \gtrsim 10^7$ GeV 
as sufficient for successful leptogenesis. In addition we also found that 
while the contributions to the EDM of charged leptons are negligibly small, 
the contributions to CLFV processes are close to the sensitivities 
of present and future experiments.

\begin{acknowledgments}
R.R. would like to thank Hiranmaya Mishra for useful discussions.
C.S.F. is supported by Funda\c{c}\~ao de Amparo \`a Pesquisa do
Estado de S\~ao Paulo (FAPESP). He would also like to thank the 
%hospitality of 
Universidad de Antioquia for their hospitality
while this work was being completed.
A.D. likes to thank Council of Scientific and Industrial Research, 
Government of India, for financial support through Senior Research
Fellowship.

\end{acknowledgments}

\appendix

\section{Thermal corrections} 
\label{app:therm}

The thermal averaged reaction density is defined as

\be
\gamma\left(ab...\rightarrow ij...\right)
\equiv \Lambda_{ab...}^{ij...}\left|\mathcal{M}
\left(ab...\rightarrow ij...\right)\right|^{2} 
\, f_{a}^{\rm eq}f_{b}^{\rm eq}...
\left(1+\eta_{i}f_{i}^{\rm eq}\right)\left(1+\eta_{j}f_{j}^{\rm eq}\right)...\,,
\label{eq:therm_ave_rate}
\ee
where ${\cal M}(ab... \to ij...)$ is the amplitude for the process
$ab ... \to ij ...$ at finite temperature,
$f_i^{\rm eq} = (e^{E_i/T} - \eta_i)^{-1}$
with $\eta_i = \pm$ for $i$ representing the boson or fermion, respectively, and
\bea
\Lambda_{ab...}^{ij...} & \equiv & \int d\Pi_{a}d\Pi_{b}...
d\Pi_{i}d\Pi_{j}...\left(2\pi\right)^{4}\delta^{\left(4\right)}
\left(p_{a}+p_{b}+...-p_{i}-p_{j}-...\right), \nonumber \\
d\Pi_{i} & \equiv & \frac{d^{3}p_{i}}{\left(2\pi\right)^{3}2E_{i}}.
\label{eq:gen_def0}
\eea

\begin{figure}
\includegraphics[scale=0.7]{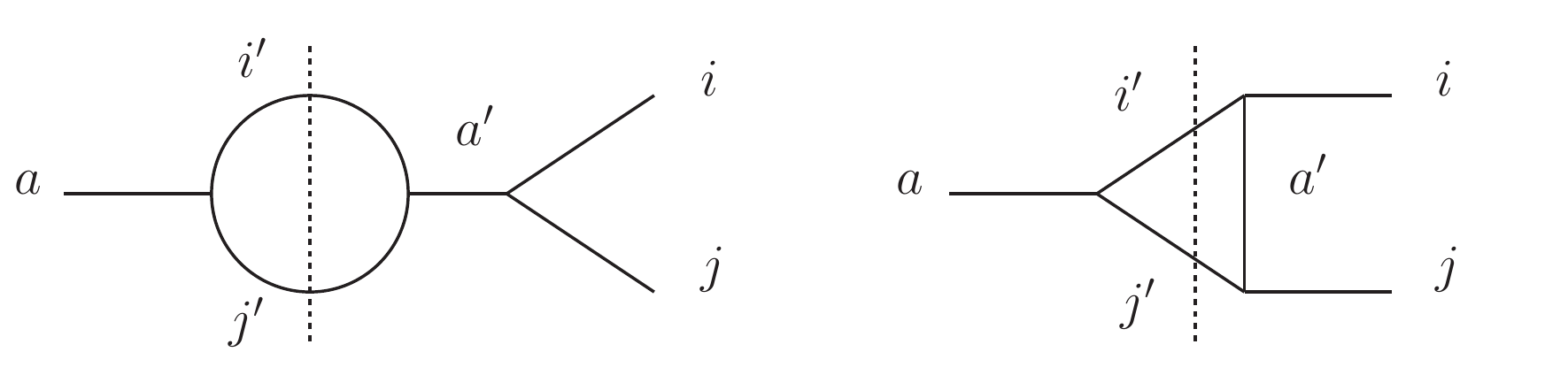}
\caption{One-loop diagrams for decay $a\to ij$\label{fig:1-loop-diagrams}.}
\end{figure}

The $CP$ asymmetry for the decay $a\to ij$ 
which arises from the interferences between tree-level and one-loop diagrams 
as shown in Fig. \ref{fig:1-loop-diagrams} is defined as

\begin{eqnarray}
\epsilon & \equiv & \frac{\gamma\left(a\to ij\right)-\gamma\left(a\to\overline{ij}\right)}{\sum_{k,l}\left[\gamma\left(a\to kl\right)+\gamma\left(a\to\overline{kl}\right)\right]}\nonumber \\
 & = & \frac{\int d\Pi_{a}f_{a}^{{\rm eq}}\int d\Phi_{ij}\left(1+\eta_{i}f_{i}^{{\rm eq}}\right)\left(1+\eta_{j}f_{j}^{{\rm eq}}\right)\left[\left|{\cal M}\left(a\to ij\right)\right|^{2}-\left|{\cal M}\left(a\to\overline{ij}\right)\right|^{2}\right]}{\sum_{k,l}\int d\Pi_{a}f_{a}^{{\rm eq}}\int d\Phi_{kl}\left(1+\eta_{k}f_{k}^{{\rm eq}}\right)\left(1+\eta_{l}f_{l}^{{\rm eq}}\right)\left[\left|{\cal M}\left(a\to kl\right)\right|^{2}+\left|{\cal M}\left(a\to\overline{kl}\right)\right|^{2}\right]},
\label{eq:cpasym_finiteT}
\end{eqnarray}
where the two-body phase space integral is
\begin{equation}
\int d\Phi_{ij}=\int d\Pi_{i}d\Pi_{j}\left(2\pi\right)^{4}\delta^{(4)}\left(p_{a}-p_{i}-p_{j}\right).
\end{equation}
Ignoring the thermal motion of $a$ with respect to the thermal bath,
i.e. setting $E_{a}=M_{a}$, we can drop the integral $\int d\Pi_{a}$ in
both the numerator and denominator. In this case the phase space integral
can be carried out analytically. In order to obtain the thermal factors 
associated with the intermediate on-shell states, one necessarily 
needs to calculate the amplitudes in Eq. \eqref{eq:cpasym_finiteT} 
using thermal field theory. In the real-time formalism of thermal field theory, 
one needs to double the number of degrees of freedom (introducing type-1 and type-2 fields) 
resulting in a $2\times 2$ matrix structure for the thermal propagator (see, e.g., \cite{Giudice:2003jh}). 
However, at one loop, we can take all the vertices connected to external legs to be of type 1
and, hence, we only need to consider the (11) element of the thermal propagator.
The (11) component for the boson propagator is \cite{Giudice:2003jh}
\be
D_B^{11} = \frac{i}{p^2 - m_B(T)^2 + i\epsilon} + 2 \pi f_B(|p_0|) \delta(p^2 - m_B(T)^2),
\label{eq:boson_prog}
\ee
where $m_B(T)$ is the boson thermal mass, and the cut propagators are
\be
D_B^{\pm} = 2 \pi \left[\theta(\pm p_0) + f_B(|p_0|) \right]\delta(p^2 - m_B(T)^2).
\label{eq:boson_prog_cut}
\ee
For fermions, the structure of the propagator is more involved \cite{Giudice:2003jh}.
For simplicity, we approximate the (11) fermion propagator by 
\be
D_F^{11} = \slashed{p}\left[\frac{i}{p^2 - m_F(T)^2 + i\epsilon} - 2 \pi f_F(|p_0|) \delta(p^2 - m_F(T)^2)\right],
\label{eq:fermion_prog}
\ee
where $m_F(T)$ is the fermion thermal mass and the cut propagators are
\be
D_F^{\pm} = 2 \pi  \slashed{p}\left[\theta(\pm p_0) - f_F(|p_0|) \right]\delta(p^2 - m_F(T)^2).
\label{eq:fermion_prog_cut}
\ee
In Eqs. \eqref{eq:fermion_prog} and \eqref{eq:fermion_prog_cut},
the propagators $\sim \slashed{p}$ are without a mass term as the bare fermion mass is zero, 
and the thermal mass does not have chiral properties.
Also, as implicit in the propagators,  
we have considered the dispersion relation as that of a free particle with a 
thermal mass, instead of the actual 
dispersion relation including thermal corrections. Although this is an 
underestimate of the actual dispersion relation, 
the error is within 10\% \cite{Weldon:1982bn}.
The above also implies that in Eq.~\eqref{eq:fermion_prog} 
we have ignored the fact that due to the interactions with the thermal bath 
the two poles of the fermion propagator have different dispersion relations 
which can lead to an order of magnitude correction to leptogenesis in the weak washout regime 
and an order of 1 correction in the strong washout 
regime \cite{Kiessig:2010pr,Kiessig:2011fw,Kiessig:2011ga}.

Keeping the above caveats in mind and applying finite temperature 
``cutting rules'' (more discussion below), we obtain
\begin{eqnarray}
\label{eq:cp_asym_sim}
\epsilon & \simeq & 
\frac{\sum_{i',j'}\left[\left|{\cal M}^{0}\left(a\to ij\right)\right|^{2}
-\left|{\cal M}^{0}\left(a\to\overline{ij}\right)\right|^{2}\right]r_{ai'j'}(T)c_{aij}(T)}
{\sum_{k,l}\left[\left|{\cal M}^{0}\left(a\to kl\right)\right|^{2}
+\left|{\cal M}^{0}\left(a\to\overline{kl}\right)\right|^{2}\right]c_{akl}(T)},
\end{eqnarray}
where ${\cal M}^{0}\left(a\to ij\right)$ is the amplitude for $a\to ij$
at zero temperature and the sum over $i'j'$ in the numerator is over
intermediate states in the loop which go on shell as shown as the
``cuts'' in Fig. \ref{fig:1-loop-diagrams}. In Eq.~\eqref{eq:cp_asym_sim} 
the $r_{aij}(T)$'s are the thermal factors 
associated with the on-shell intermediate states,
while the $c_{aij}(T)$'s are those associated with the final states. 
In the case of self-energy contributions, the factorized form as a product of
thermal-dependent and zero temperature terms as in Eq.~\eqref{eq:cp_asym_sim} 
always holds (under the approximation that $a$ is at rest with respect to the thermal bath) 
while in the case of vertex diagrams, further approximations are required.
One approximation we have made is to factorize out the temperature-dependent 
terms including the kinematic factors, and then to 
set the thermal masses in the rest of the terms to zero, 
which gives us expressions for these terms that coincide with the
zero temperature results. In addition, we ignore the contributions 
from the cuts through $a'$ and $i'$, or $a'$ and $j'$ in the vertex diagrams, which 
as shown in Ref.~\cite{Garbrecht:2010sz} in non-SUSY type-I leptogenesis can 
give corrections depending on the $a-a'$ mass ratio, 
for example, at the level of 10\% for $m_{a'}/m_a=1.1$.
Under these approximations, the temperature-dependent terms for both self-energy and 
vertex diagrams are the same and are given by
\begin{eqnarray}
c_{aij}(T) & = & \left[1+\eta_{a}\left(1-\delta_{bi}\delta_{bj}\right)\left(\eta_{i}x_{i}+\eta_{j}x_{j}\right)\right]\lambda\left(1,x_{i},x_{j}\right)\left(1+\eta_{i}f_{i}^{{\rm eq}}\right)\left(1+\eta_{j}f_{j}^{{\rm eq}}\right),\label{eq:final_states}\\
r_{aij}(T) & = & \left[1+\eta_{a}\left(1-\delta_{bi}\delta_{bj}\right)\left(\eta_{i}x_{i}+\eta_{j}x_{j}\right)\right]\lambda\left(1,x_{i},x_{j}\right)\left(1+\eta_{i}f_{i}^{{\rm eq}}+\eta_{j}f_{j}^{{\rm eq}}\right),\label{eq:int_states}
\end{eqnarray}
with $\delta_{bi}=1\left(0\right)$ if $i$ is a boson (fermion) and 
\begin{eqnarray}
\lambda(1,x,y) & = & \sqrt{\left(1+x-y\right)^{2}-4x},\;\;\;\;\;\;\;\;
x_{i} = \frac{m_{i}\left(T\right)^{2}}{M_a^{2}},\nonumber \\
E_{i} & = & \frac{M_{a}}{2}\left(1+x_{i}-x_{j}\right),\;\;\;\;\;\;\;\;
E_{j} = M_{a}-E_{i}=\frac{M_{a}}{2}\left(1-x_{i}+x_{j}\right).
\end{eqnarray}
For the statistical factors in $r_{aij}(T)$, we applied the finite temperature
cutting rules by considering causal (i.e., retarded or advanced) n-point
functions as pointed out by Ref.~\cite{Garny:2010nj} which gives 
the dependence on the distribution functions $1+\eta_{i}f_{i}^{{\rm eq}}+\eta_{j}f_{j}^{{\rm eq}}$
in agreement with the results derived from nonequilibrium quantum
field theory \cite{Garbrecht:2010sz,Buchmuller:2000nd,De Simone:2007rw,Garny:2009rv,Garny:2009qn,Garny:2010nz,
Garny:2011hg,Frossard:2012pc,Anisimov:2010aq,Anisimov:2010dk,Beneke:2010wd,Beneke:2010dz,
Garbrecht:2011aw,Dev:2014wsa} in contrast to the results of Refs. \cite{Covi:1997dr,Giudice:2003jh} which obtained
$1+\eta_{i}f_{i}^{{\rm eq}}+\eta_{j}f_{j}^{{\rm eq}}+\eta_{i}\eta_{j}f_{i}^{{\rm eq}}f_{j}^{{\rm eq}}$ 
when time-ordered n-point functions are considered instead. 
Notice that the imaginary time formalism also gives the statistical factors in agreement with 
the result of nonequilibrium quantum field theory \cite{Kiessig:2010pr,Kiessig:2011fw,Kiessig:2011ga}.

Now we can apply the general results \eqref{eq:final_states} and \eqref{eq:int_states} 
to soft leptogenesis. For the decays $\widetilde{N}_{\pm}\to\ell_{\alpha}\widetilde{H}_{u}$
and $\widetilde{N}_{\pm}\to\widetilde{\ell}H_{u}$, the relevant thermal
factors are obtained from Eqs.~\eqref{eq:final_states} and \eqref{eq:int_states}
to be
\begin{eqnarray}
c_{F}(T) & = & \left(1-x_{\ell}-x_{\widetilde{H}_{u}}\right)\lambda\left(1,x_{\ell},x_{\widetilde{H}_{u}}\right)\left(1-f_{\ell}^{{\rm eq}}\right)\left(1-f_{\widetilde{H}_{u}}^{{\rm eq}}\right),\nonumber \\
c_{B}(T) & = & \lambda\left(1,x_{\widetilde{\ell}},x_{H_{u}}\right)\left(1+f_{\widetilde{\ell}}^{{\rm eq}}\right)\left(1+f_{H_{u}}^{{\rm eq}}\right),\nonumber \\
r_{F}(T) & = & \left(1-x_{\ell}-x_{\widetilde{H}_{u}}\right)\lambda\left(1,x_{\ell},x_{\widetilde{H}_{u}}\right)\left(1-f_{\ell}^{{\rm eq}}-f_{\widetilde{H}_{u}}^{{\rm eq}}\right),\nonumber \\
r_{B}(T) & = & \lambda\left(1,x_{\widetilde{\ell}},x_{H_{u}}\right)\left(1+f_{\widetilde{\ell}}^{{\rm eq}}+f_{H_{u}}^{{\rm eq}}\right).
\end{eqnarray}
The relevant thermal masses are \cite{Giudice:2003jh}
\begin{eqnarray}
m_{\widetilde{\ell}}(T)^{2} & = & 2m_{\ell}(T)^{2} 
= \left(\frac{3}{8}g_{2}^{2}+\frac{1}{8}g_{Y}^{2}\right)T^{2},\nonumber \\
m_{H_{u}}(T)^{2} & = & 2m_{\widetilde{H}_{u}}(T)^{2}
=\left(\frac{3}{8}g_{2}^{2}+\frac{1}{8}g_{Y}^{2}+\frac{3}{4}\lambda_{t}^{2}\right)T^{2}.
\end{eqnarray}

Next we prove a useful identity (in the context of the 
approximations made above)
as follows
\begin{eqnarray}
r_{F}(T)c_{B}(T)-r_{B}(T)c_{F}(T) 
& \propto & \left(1-f_{\ell}^{{\rm eq}}-f_{\widetilde{H}_{u}}^{{\rm eq}}\right)\left(1+f_{\widetilde{\ell}}^{{\rm eq}}\right)\left(1+f_{H_{u}}^{{\rm eq}}\right)-\left(1+f_{\widetilde{\ell}}^{{\rm eq}}+f_{H_{u}}^{{\rm eq}}\right)\left(1-f_{\ell}^{{\rm eq}}\right)\left(1-f_{\widetilde{H}_{u}}^{{\rm eq}}\right)\nonumber \\
 & = & \left(e^{E_{\widetilde{N}}/T}-1\right)f_{\ell}^{{\rm eq}}f_{\widetilde{H}_{u}}^{{\rm eq}}e^{E_{\widetilde{N}}/T}f_{\widetilde{\ell}}^{{\rm eq}}f_{H_{u}}^{{\rm eq}}-\left(e^{E_{\widetilde{N}}/T}-1\right)f_{\widetilde{\ell}}^{{\rm eq}}f_{H_{u}}^{{\rm eq}}e^{E_{\widetilde{N}}/T}f_{\ell}^{{\rm eq}}f_{\widetilde{H}_{u}}^{{\rm eq}}\nonumber \\
 & = & 0.\label{eq:RC_identity}
\end{eqnarray}
In the second line above, we have made use of the following identity
\begin{eqnarray}
\left(1+\eta_{i}f_{i}^{{\rm eq}}\right)\left(1+\eta_{j}f_{j}^{{\rm eq}}\right) & = & e^{\left(E_{i}+E_{j}\right)/T}f_{i}^{{\rm eq}}f_{j}^{{\rm eq}},
\end{eqnarray}
and the conservation of energy $E_{\widetilde{N}}=E_{\ell}
+E_{\widetilde{H}_{u}}=E_{\widetilde{\ell}}+E_{H_{u}}$.
Notice that this identity also holds if instead of using the factor 
$1+\eta_{i}f_{i}^{{\rm eq}}+\eta_{j}f_{j}^{{\rm eq}}$
in $r_{aij}(T)$, one uses $1+\eta_{i}f_{i}^{{\rm eq}}
+\eta_{j}f_{j}^{{\rm eq}}+\eta_{i}\eta_{j}f_{i}^{{\rm eq}}f_{j}^{{\rm eq}}$
as obtained in Refs. \cite{Covi:1997dr,Giudice:2003jh}.

Finally, it can be shown that the $CP$ asymmetries from gaugino 
contributions \cite{Grossman:2004dz,Fong:2009iu}
for the decays of $\widetilde N_\pm$ to scalars and fermions are, respectively, 
given by $\epsilon_g r_F(T)c_B(T)$ and $-\epsilon_g r_B(T)c_F(T)$ where 
$\epsilon_g$ is some temperature-independent term. 
Using the identity (\ref{eq:RC_identity}), these contributions sum 
up to zero illustrating the cancellations
pointed out by Ref.~\cite{Garbrecht:2013iga}.

\section{Special cases of mixing $CP$ asymmetries} 
\label{app:CP_asym}

Here we will discuss two specific cases of mixing CP asymmetries.

\begin{itemize}
\item[(a)] Universal trilinear scenario: $A_\alpha = A Y_\alpha$.

This is the scenario considered in Refs. \cite{D'Ambrosio:2003wy,Grossman:2003jv,Fong:2008mu}.
In this scenario, we are always in the regime (i) $Y_\alpha \gg |A_\alpha|/M$, 
and from Eq.~\eqref{eq:ep_S_tot_i}, we obtain \cite{Fong:2008mu}
\bea
\epsilon^S_{\pm \alpha}
& \simeq & P_\alpha\, \bar\epsilon\, \frac{c_F(T) - c_B(T)}{c_F(T) + c_B(T)} r_B(T),
%& & + \,\frac{1}{4\pi} 
%P_\alpha \sum_\beta \frac{{\rm Im} (A)}{M} 
%\frac{4 B \Gamma_Y}{4 B^2 + \Gamma_Y^2} \frac{\delta_\pm c_F(T)}{(1+\delta_\pm)c_F(T) + c_B(T)} r_B(T),
\label{eq:ep_S_UTS}
\eea
with
\begin{eqnarray}
\bar{\epsilon} & \equiv & \frac{{\rm Im}\left(A\right)}{M}\frac{4B\Gamma_{Y}}{4B^{2}+\Gamma_{Y}^{2}}.
\end{eqnarray}
%\bea
%\epsilon^S_{\alpha}
%& \simeq & - P_\alpha \frac{{\rm Im} (A)}{M}
%R_b(T) \frac{4 B \Gamma_Y}{4 B^2 + \Gamma_Y^2}\,
%\frac{c_B(T) - c_F(T)}{c_B(T) + c_F(T)}.    %\nonumber \\
%& & + \, P_\alpha \frac{|A|^2}{M^2} \frac{{\rm Im} (A)}{M} 
%\frac{4 B \Gamma}{4 B^2 + \Gamma^2}\, \frac{R_b(T) C_b(T)}{C_f(T) + C_b(T)},
%\label{eq:ep_S_UTS}
%\eea
%The CP asymmetry is vanishing in the limit of zero temperature
%where $c_{B,F}(0) = 1$.

\item[(b)] Simplified misaligned scenario: $A_\alpha = A Y^2/(3Y_\alpha)$.

This is a specific scenario considered in Ref.~\cite{Fong:2010zu}.
In this scenario, we have from Eq.~\eqref{eq:G_T}
\be
G_\pm(T) \simeq Y^2 \left[c_F(T) + c_B(T) \right] 
+ Y^2 \frac{|A|^2}{M^2} d \, c_B(T),
\label{eq:G_SMS}
\ee
%\be
%G_\pm(T) =  Y^2\left[c_B(T) + c_F(T)\right] +
%Y^2\left[\frac{|A|^2}{M^2}\,\sum_\alpha\frac{1}{9P_\alpha} \right]c_B(T)
%\ee
where $d \equiv \sum_\alpha 1/(9P_\alpha) \geq 1$, and
the minimum occurs at $P_\alpha = 1/3$ for all $\alpha$.  
In Eq. \eqref{eq:G_SMS}, we have dropped terms of ${\cal O}(Y^2 m_{\rm SUSY}/M)$ 
except the second term which could dominate over the first when $|A|^2/M^2 \gg d^{-1}$.
This condition can only be fulfilled if $P_\alpha$ 
deviates significantly from $1/3$, i.e. a very hierarchical $P_\alpha$. 

First let us consider the case $|A|^2/M^2 \ll d^{-1}$. 
From Eq. \eqref{eq:ep_S_tot} we obtain
\begin{eqnarray}
\epsilon_{\pm\alpha}^{S} & \simeq & P_{\alpha}
\bar{\epsilon}\,\frac{c_{F}\left(T\right)-c_{B}\left(T\right)}{c_{F}\left(T\right)+c_{B}\left(T\right)}r_{B}\left(T\right) 
+ \frac{1}{9P_{\alpha}}\frac{\left|A\right|^{2}}{M^{2}}\bar{\epsilon}\,\frac{c_{B}\left(T\right)}{c_{F}\left(T\right)+c_{B}\left(T\right)}r_{B}\left(T\right).
\end{eqnarray}
Regarding the first ``thermal'' term, Ref.~\cite{Fong:2010zu} 
made a mistake in that the lepton flavors in the self-energy
loop were not summed over resulting in expressions which were independent
of $P_{\alpha}$. Since this term coincides with Eq. \eqref{eq:ep_S_UTS} 
there is no enhancement for the simplified misaligned scenario case compared to 
the universal trilinear scenario case as claimed 
in Ref.~\cite{Fong:2010zu}.\footnote{Nonetheless the idea of considering a generic 
$A$ term to enhance the efficiency was correct as shown in the current work.}

In Ref.~\cite{Fong:2010zu}, the second ``nonthermal'' term is ignored assuming
it is always smaller than the first one. This is always true when the
$P_{\alpha}$'s are not hierarchical or only mildly hierarchical.
The ``nonthermal'' term can dominate only in a hierarchical 
scenario when some of the $P_{\alpha}$ fulfill
\begin{eqnarray}
P_{\alpha} & \ll & \frac{1}{3}\frac{\left|A\right|}{M}.
\label{eq:SMS_cond}
\end{eqnarray}
When the condition above is fulfilled, we always have a mixed scenario
where we have both thermal and nonthermal contributions -- since
$\sum_{\alpha}P_{\alpha}=1$, some of the $P_{\alpha}$'s 
cannot fulfill Eq. \eqref{eq:SMS_cond}. 

For the case $\left|A\right|^{2}/M^{2}\gg d^{-1}$, which can only
happen when $P_{\alpha}$ is very hierarchical, we have 
from Eq. \eqref{eq:ep_S_tot}
\begin{eqnarray}
\epsilon_{\pm\alpha}^{S} & \simeq & P_{\alpha}
\frac{M^2}{\left|A\right|^{2}d}\,
\bar{\epsilon}_A\,
\frac{c_{F}\left(T\right)-c_{B}\left(T\right)}{c_{B}\left(T\right)}r_{B}
\left(T\right)
+ \frac{1}{9P_{\alpha}d}\bar{\epsilon}_A\, r_{B}\left(T\right)\,,
\label{eq:SMS2}
\end{eqnarray}
where
\begin{eqnarray}
\bar{\epsilon}_A & \equiv & 
\frac{{\rm Im}\left(A\right)}{M}
\frac{4B\Gamma_{Y}}{4B^{2}+\Gamma_{Y}^{2}
\left(\frac{\left|A\right|^{2}d}{2M^{2}}\right)^2}.
\end{eqnarray}
The condition for the second ``nonthermal'' term 
in Eq. \eqref{eq:SMS2} to dominate is again Eq. \eqref{eq:SMS_cond}.

\end{itemize}

\end{document}